\documentclass[10pt,
final,
conference,
twocolumn,
 comsoc,
letterpaper
,times,mathptm]{IEEEtran} 

\usepackage{srcltx}
\usepackage[psamsfonts]{amsfonts}
\usepackage{amsmath}
\usepackage{amssymb}
\usepackage{amsfonts}
\usepackage{graphics}
\usepackage{tikz} 
\usetikzlibrary{fit,positioning}
\usetikzlibrary{arrows,positioning,automata}
\usepackage{psfrag}
\usepackage{epsfig}
\usepackage{subfigure}
\usepackage{array}
\usepackage{algorithm}
\usepackage{algorithmic}
\usepackage{pifont}
\usepackage{srcltx}
\usepackage[psamsfonts]{amsfonts}
\usepackage{makeidx}  
\usepackage{bbm}
\usepackage{hhline}
\usepackage{eufrak}
\usepackage{yfonts}
\usepackage{color}
\usepackage{url}
\usepackage{dsfont}
\usepackage{caption}
\usepackage[square, comma, sort&compress, numbers]{natbib}
\usepackage{balance} 

\interdisplaylinepenalty=2500 \algsetup{indent=2em}
\title{Adaptive Non-uniform Compressive Sampling for Time-varying Signals
}
\author{Alireza Zaeemzadeh, Mohsen Joneidi, and Nazanin Rahnavard\\
School of Electrical Engineering and Computer Science \\
University of Central Florida \\
Emails: {\{zaeemzadeh, joneidi, and nazanin \} @eecs.ucf.edu}
}
\begin{document}
\renewcommand{\textfraction}{0}
\maketitle
\IEEEpeerreviewmaketitle

\begin{abstract}
In this paper, adaptive non-uniform compressive sampling (ANCS) of time-varying signals, which are sparse in a proper basis, is introduced. ANCS employs the measurements of previous time steps to distribute the sensing energy among coefficients more intelligently. To this aim, a Bayesian inference method is proposed that does not require any prior knowledge of importance levels of coefficients or sparsity of the signal. Our numerical simulations show that ANCS is able to achieve the desired non-uniform recovery of the signal. Moreover, if the signal is sparse in canonical basis, ANCS can reduce the number of required measurements significantly. 
\end{abstract}
\vspace{-2.5mm}
\section{Introduction}\label{sec:intro}
Compressed sensing (CS) \cite{Donoho2006CompressedSensing,Candes2006CompressiveSampling} states that most of the signals of scientific interest can be approximated very accurately using a smaller number of measurements, compared to the dimension of the signal. For that, the signal needs to be sparse or have a sparse representation in terms of proper sparsifying bases. This observation has a huge impact in signal processing, machine learning, and statistics. Mathematically speaking, the goal of the CS problem is to recover the signal $\boldsymbol{x} = [x_1,x_2,\dots, x_N]^T$ with length $N$ from its undersampled random projections, also referred to as measurements. $M$ random projections are generated using a  measurement matrix $\boldsymbol{\Phi} \in \mathbb{R}^{M\times N}$ from the linear measurement process,
$
\boldsymbol{y} = \boldsymbol{\Phi} \boldsymbol{x} + \boldsymbol{n},
$ where $\boldsymbol{y} = [y_1,y_2,\dots, y_M]^T$ represents the measurement vector and $\boldsymbol{n}$ denotes the corrupting noise. \par 

Signal $\boldsymbol{x}$ is said to be $K$-sparse if it has at most $K$ non-zero entries in a proper basis. The sparsity of $\boldsymbol{x}$ can be exploited to find a unique solution of the underdetermined system equation with high probability from $\mathcal{O}(K\log(\frac{N}{K}))$ measurements \cite{Candes2006CompressiveSampling}. \par 

In this paper, we consider the problem of reconstructing a correlated time series of such compressible vectors from their noisy undersampled measurement. Particularly, we are interested in approximating the time series $\{ \boldsymbol{x}^{(1)}, \boldsymbol{x}^{(2)}, \dots \}$ from the measurement time series $\{ \boldsymbol{y}^{(1)}, \boldsymbol{y}^{(2)}, \dots \}$. In many real-world applications, the signal of interest has a substantial correlation in time. The main idea is to incorporate the knowledge from the previous estimates of the signal to achieve a more accurate estimation of the signal at the current time step. \par 

Moreover, in many applications, different parts of the signal have different recovery requirements. Thus, different coefficients of the signal have different \emph{importance levels}. For instance, in video processing, it is desired to recover the salient area more accurately. Moreover, if the signal is sparse in canonical basis, we are interested in reconstructing the large coefficients with less error. \emph{Non-uniform} acquisition and recovery of signal is desirable in many applications such as image processing \cite{shahrasbi2016model}, camera sensor networks \cite{Uddin2011PhotoNet:Awareness,Rahimpour2016DistributedNetworks}, wireless sensor networks \cite{Leinonen2015SequentialNetworks}, collaborative vector estimation \cite{Sani2016DistributedNetworks}, component analysis \cite{Rahmani2016ASampling,Hosseini2016Cloud-basedPrediction}, and internet of things \cite{Fragkiadakis2014AdaptiveApplications}.

In this work, we propose an adaptive framework to design a non-uniform measurement matrix, which contrast with \emph{dynamic CS algorithms} \cite{Ziniel2013DynamicPassing,Wijewardhana2016AMeasurements,Shahrasbi2011TC-CSBP:Propagation} focusing only on the recovery step. Our method is also distinct from the \emph{adaptive CS} \cite{Malloy2014Near-OptimalSensing,Braun2015Info-GreedySensing,Haupt2012SequentiallySensing} methods that are concerned with reconstructing signals, which are static over time. Here, similar to adaptive CS, the main idea is to concentrate the sensing energy on the more important coefficients, by designing a proper measurement matrix. However, due to dynamic nature of the problem, the algorithm should not make firm decisions about the location of more important coefficients. Hence, soft importance level information is advantageous. To infer the importance level of each coefficient at each time step, a generative model is imposed on the coefficients and the parameters of the model are updated in an online fashion. \par

\begin{figure}
\centering
\includegraphics[width=\columnwidth,angle=0]{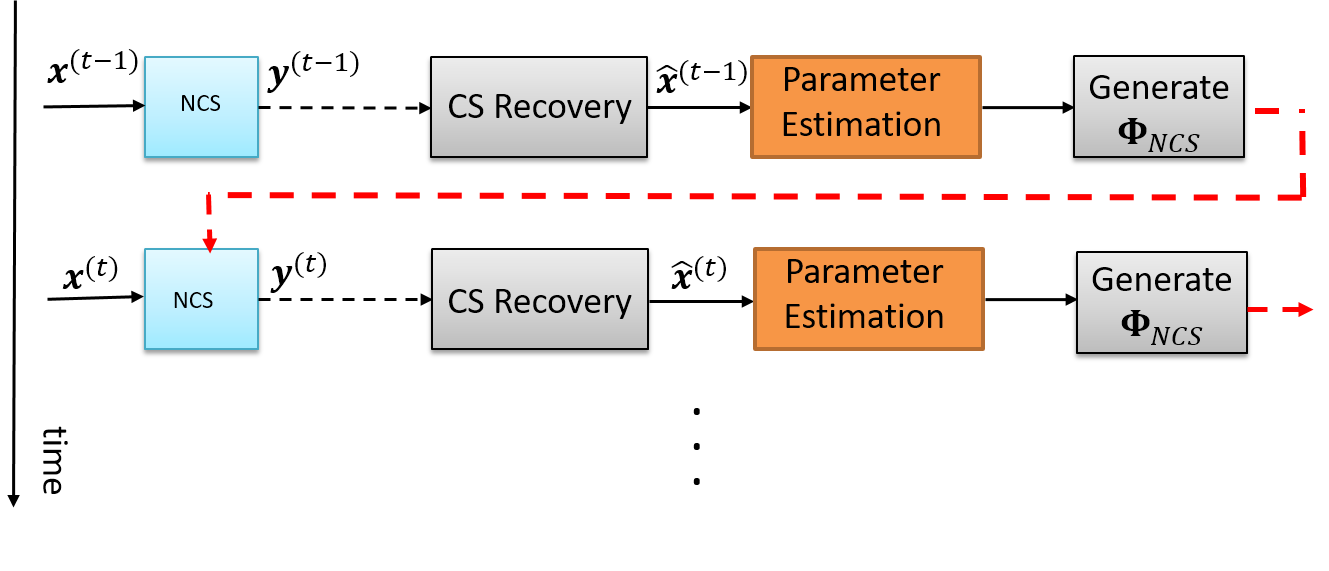}\vspace{-1.5mm}
\caption{\small Overall block diagram of the proposed framework.Reconstructed signal, at each time step, is utilized to generate the measurement matrix. }
\label{fig:block_diagram}
\vspace{-5mm}
\end{figure}

Figure \ref{fig:block_diagram} shows the overall architecture of the proposed method. At each time step, after reconstructing the signal, by using a conventional CS recovery algorithm, the importance levels of the coefficients are inferred mathematically. The importance levels are further employed to design the measurement matrix for the next time step. \par 

The rest of this paper is organized as follows. In Section \ref{sec:system_model}, the system model is presented. Then, the generative model of the proposed Bayesian framework is introduced in Section \ref{sec:Bayesian}. In Section \ref{sec:matrix_design}, the inferred importance level information are used to design the measurement matrix for sensing. Finally, Section \ref{sec:results} presents the simulation results and Section \ref{sec:conclusions} draws conclusions. \par 


\section{System Model}\label{sec:system_model}
We consider recovery of a vector-valued time series $\{ \boldsymbol{x}^{(1)}, \boldsymbol{x}^{(2)}, \dots \}$ from the linear measurements given by 
\begin{equation}
\label{eq:meas_model}
\boldsymbol{y}^{(t)} = \boldsymbol{\Phi}^{(t)}\boldsymbol{x}^{(t)} + \boldsymbol{n}^{(t)}, \qquad t = 1,2,\dots
\end{equation}
where $\boldsymbol{n}^{(t)} \in \mathbb{R}^M$ represents the noise and is modeled as an additive white Gaussian noise (AWGN) with $\boldsymbol{n}^{(t)} \sim \mathcal{N}(\boldsymbol{0},\sigma_{n}^2\boldsymbol{I}_{M})$. \par 

It is assumed that the signal of interest $\boldsymbol{x}^{(t)}$ is compressible and contains coefficients with different importance levels, which are not known a priori. In many scenarios, it is desirable to have non-uniform recovery performance on different parts of signal. More important coefficients may correspond to support of a sparse vector or the salient area in a video frame. \par 

We also assume that, at each time step, using the estimation of the signal $\hat{\boldsymbol{x}}^{(t)}$, more important coefficients are tagged using a possibly erroneous algorithm. The variable $\boldsymbol{\alpha}^{(t)}$ marks the detected \emph{region of interest} (ROI) in the signal at time $t$. Specifically, $\alpha_n^{(t)} = 1$, if the $n^{\text{th}}$ coefficient of the signal is detected to be in the ROI, and $\alpha_n^{(t)} = 0$ otherwise. However, due to sensing failure, error in recovery of $\hat{\boldsymbol{x}}^{(t)}$, and/or misdetection of the ROI, $\boldsymbol{\alpha}^{(t)}$ may contain erroneous elements. \par 

As mentioned earlier, the signal of interest often exhibits substantial temporal correlation. Here, we assume that ROI, and therefore the support of non-zero entries in $\boldsymbol{\alpha}^{(t)}$, changes slowly in time. Our goal is to employ the temporal correlation to infer reliable importance level information and employ the importance levels to design a non-uniform measurement matrix.   

\section{Bayesian Inference of Importance Levels} \label{sec:Bayesian}
To extract reliable information from possibly faulty ROI data $\boldsymbol{\alpha}^{(t)}$ , we propose to employ Bayesian inference. In Bayesian framework, the goal is to infer the probability distribution of hidden variables given the observations. The hidden variables are often the parameters that are desired to be estimated. Specifically, in our model, the following hidden variables are introduced: 
\begin{enumerate}
\item Coefficient-specific reliability $u_{n} \in \{ 0,1 \}$, which is either $0$ or $1$ and describes the reliability of ROI data of the $n^{\text{th}}$ coefficient.
\item Overall reliability $r \in [0,1]$, denoting the overall trustworthiness of the ROI detection algorithm. For small values of $r$, the algorithm is more prone to reporting faulty data. A generally reliable algorithm will report trustworthy measurements on most of the coefficients.
\item Importance level for each coefficient $c_{n} \in [0,1]$, describing the probability that coefficient $n$ is in ROI. 
\end{enumerate}

As mentioned earlier, in the proposed generative model, $\alpha_n$ is the observed variable. If $\alpha_n = 1$, the $n^{\text{th}}$ coefficient is detected to be in ROI, and $\alpha_n = 0$ otherwise. In this model, coefficient-specific reliability and overall reliability model the faulty data. Without them, all the observations would be assumed to be trustworthy, which is not the case in real-world scenarios. \par 

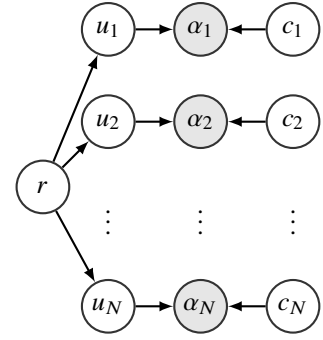
\begin{figure}
\centering
\scalebox{1}{
\begin{tikzpicture}
\tikzstyle{main}=[circle, minimum size = 7mm, thick, draw =black!80, node distance = 5mm,inner sep=1pt]
\tikzstyle{connect}=[-latex, thick]
\tikzstyle{box}=[rectangle, draw=black!100]
  \node[main, fill = black!10] (alpha1) [] {$\alpha_{1}$ };
  \node[main, fill = black!10] (alpha2) [below=of alpha1] {$\alpha_{2}$ };
  \node[main] (alphadots) [below=of alpha2,draw=none] {$\vdots$ };
  \node[main, fill = black!10] (alphaN) [below=of alphadots] {$\alpha_{N}$ };
  \node[main] (u1) [left= of alpha1] {$u_{1}$ };
    \node[main] (u2) [below=of u1] {$u_{2}$ };
  \node[main] (udots) [below=of u2,draw=none] {$\vdots$ };
  \node[main] (uN) [below=of udots] {$u_{N}$ };
  \node[main] (c1) [right= of alpha1] {$c_{1}$ };
    \node[main] (c2) [below=of c1] {$c_{2}$ };
  \node[main] (cdots) [below=of c2,draw=none] {$\vdots$ };
  \node[main] (cN) [below=of cdots] {$c_{N}$ };
  \node[main] (r) [ below left=of u2] {$r$ };
   \path (u1) edge [connect] (alpha1)
	     (u2) edge [connect] (alpha2)
	     (uN) edge [connect] (alphaN)
         (c1) edge [connect] (alpha1)
	     (c2) edge [connect] (alpha2)
	     (cN) edge [connect] (alphaN)
         (r) edge [connect] (u1)
	     (r) edge [connect] (u2)
	     (r) edge [connect] (uN)    
         ;
\end{tikzpicture}
}
\caption{\small Graphical representation of the generative model. }
\label{fig:graph_model}
\vspace{-4mm}
\end{figure}

Figure \ref{fig:graph_model} illustrates the graphical representation of the proposed generative model. The arrows in the graph represent the dependency among the variables. Hence, the observed ROI data depends on the actual importance level of the coefficients and the reliability of algorithm in detecting the ROI coefficients. 
The goal of the inference algorithm  is to obtain the probability distribution of the overall reliability, coefficient-specific reliability, and the importance levels, given the ROI data. At each time step $t$, the proposed model can be formulated as follows, for $n=1, \dots, N$:
\begin{equation}
\begin{aligned}
 r &\sim ~\operatorname{Beta}(b^{1},b^{0})  \\
 u_{n} &\sim ~\operatorname{Bernoulli}(r)  \\
 c_{n} &\sim ~\operatorname{Beta}(\beta^{1}_{n},\beta^{0}_{n})  \\
 \alpha_{n}^{(t)} &\sim ~ u_{n} \operatorname{Bernoulli}(c_{n}) +  (1- u_{n}) \operatorname{Bernoulli}(1 - c_{n})
\label{eq:gen_model}
\end{aligned}
\end{equation}

The observed variable $\alpha_n^{(t)}$ is modeled with summation of two Bernoulli distributions. This means that if the ROI data for $n^{\text{th}}$ coefficient is reliable, i.e. $u_n = 1$, $\alpha_n$ will be sampled from a Bernoulli distribution with true parameter for importance level, i.e. $c_n$. Otherwise, it will be sampled from $\operatorname{Bernoulli}( 1 - c_{n})$ and will be more probable to report faulty data. Since $c_n$ is used as the parameter of a Bernoulli distribution, it is the natural choice to model it with a Beta distribution. This is due to the fact that the conjugate prior for Bernoulli distribution is Beta distribution. \par 

Similarly, the variable representing the overall reliability, i.e. $r$,  is modeled with a Beta distribution. This is because the coefficient-specific reliability variables are sampled from $\operatorname{Bernoulli}(r)$. This means that if the ROI detection is reliable in general, ROI data on most of the coefficients will be reliable. This prior links the performance of the algorithm on different coefficients and reduces the chance of overfitting the coefficient-specific reliability. \par 

As mentioned earlier, the goal of the inference algorithm is to obtain the distribution of hidden variables, given the observations, i.e. $\mathbb{P} \{\boldsymbol{c},\boldsymbol{u}, r| \mathcal{A} \}$. For compactness of notation, we set $\boldsymbol{c} = \{c_1,c_2,\dots, c_N\}$, $\boldsymbol{u} = \{u_1,u_2,\dots, u_N\}$, and $\mathcal{A} = \{ \boldsymbol{\alpha}^{(1)},  \boldsymbol{\alpha}^{(2)},\dots \}$. At each time step, after receiving the ROI data, the distribution of hidden variables are inferred by exploiting the data and the prior belief, represented by the prior distribution $\mathbb{P} \{\boldsymbol{c},\boldsymbol{u}, r\}$. 

For that, we need to specify the joint distribution of the observation and the hidden variables. Specifically, using the model formulated in (\ref{eq:gen_model}), we have:
\begin{equation}
\label{eq:joint_dist}
\begin{aligned}
\mathbb{P}\{\mathcal{A} ,\boldsymbol{u},\boldsymbol{c},r \} =
\prod_{t = 1}^{\infty}\prod_{n = 1}^{N}
&\mathbb{P}\{\alpha_{n}^{(t)}|u_{n},c_{n}\}
\mathbb{P}\{u_{n}|r\}\\ 
&\mathbb{P}\{c_{n}|\beta^1_n,\beta^0_n\}\mathbb{P}\{r|b^1,b^0\}
\end{aligned}
\end{equation}

However, due to obvious practical reasons and to limit the history of the inference, the inference is performed using a few of recent observations. For that, a sliding window of length $W$ is utilized and the parameters of the posterior distributions are inferred using only the last $W$ observations. \par

To infer the importance level of the coefficients as well as the reliability of the ROI data, we need to find the posterior distribution given the ROI data, i.e., $\mathbb{P}\{\boldsymbol{u},\boldsymbol{c},r | \mathcal{A}\}$. However, directly obtaining the posterior distributions is not computationally feasible and results in explosive number of probability factors growing exponentially with number of coefficients. To handle the intractable integrals of the inference procedure, \emph{variational inference} is often employed \cite{Jordan2008graphical,jordan1999variational,Babagholami-Mohamadabadi2014AEstimation}. \par 

In variational inference, the posterior distribution is assumed to be fully factorized over all the hidden variables. In other words, the posterior distribution is being approximated by a family of distributions, for which the inference procedure is tractable. For our model, the fully factorized approximation of the posterior distribution, also referred to as the variational distribution, is defined as:
\begin{equation}
\label{eq:variational_dist}
\mathbb{Q}\{\boldsymbol{c},\boldsymbol{u},r\} =
\prod_{n}\mathbb{Q}\{c_{n}|\hat{\beta}^1_{n},\hat{\beta}^0_{n}\}
\mathbb{Q}\{u_{n}|\tau_{n}\}\\
\mathbb{Q}\{r_{n}|\hat{b}^1,\hat{b}^0\}.
\end{equation}
where  $\hat{b}^1$, $\hat{b}^0$, $\hat{\beta}^1_{n}$, $\hat{\beta}^0_{n}$, and $\tau_{n}$ are the parameters of the factorized  distributions. By introducing the variable $\tau_{n}$,  we are seeking the best approximate of $\mathbb{P}\{\boldsymbol{u},\boldsymbol{c},r | \mathcal{A}\}$ among all the distributions $\mathbb{Q}\{\boldsymbol{c},\boldsymbol{u},r\}$, by factorizing the distribution over \emph{disjoint} groups of hidden variables. $\boldsymbol{u}$, $\boldsymbol{c}$, and $r$. 
It is worthwhile to mention that we make no further assumption about the distributions and their functional forms.\par 

 Specifically, we aim to find the best set of distributions and parameters that maximizes the lower bound of log likelihood of the observations \cite{jordan1999variational,bishop2006pattern}. The lower bound of log-likelihood of the observations can be written as \cite[Chapter 10]{bishop2006pattern}:
\begin{equation}
\label{eq:lower_bound}
\begin{aligned}
\ln ( \mathbb{P} \{ \mathcal{A}\} ) \geq
&  \int \mathbb{Q}\{\boldsymbol{c},\boldsymbol{u},r\} \ln (\mathbb{P}\{\mathcal{A},\boldsymbol{c},\boldsymbol{u},r\}) - \\
& \quad \int \mathbb{Q}\{\boldsymbol{c},\boldsymbol{u},r\} \ln (\mathbb{Q}\{\boldsymbol{c},\boldsymbol{u},r\})\\
& = \mathbb{E}\{ \ln (\mathbb{P}\{ \mathcal{A}, \boldsymbol{c},\boldsymbol{u},r\}) \} \\
& - \mathbb{E}\{ \ln (\mathbb{Q}\{\boldsymbol{c},\boldsymbol{u},r\}) \} \\
& \triangleq \mathcal{L}(\mathbb{Q}\{\boldsymbol{c},\boldsymbol{u},r\}),
\end{aligned}
\end{equation}
where the expected value is with respect to variational distribution. Hence, the problem boils down to maximizing $\mathcal{L}(\mathbb{Q}\{\boldsymbol{c},\boldsymbol{u},r\})$ to find the best variational distributions. Since the lower bound is concave with respect to each of the factorized distributions, i.e., $\mathbb{Q}\{c_{n}|\hat{\beta}^1_{n},\hat{\beta}^0_{n}\}$,$
\mathbb{Q}\{u_{n}|\tau_{n}\}$,and $\mathbb{Q}\{r_{n}|\hat{b}^1,\hat{b}^0\}$, we can determine the best approximate distributions by maximizing $\mathcal{L}(\mathbb{Q}\{\boldsymbol{c},\boldsymbol{u},r\})$ with respect to one factor at a time \cite{bishop2006pattern}. Thus, at each step, the lower bound is maximized over one factor, keeping all the other distributions. This procedure is repeated until convergence. \par 

For simplicity of notation, let us denote the whole set of hidden variables with $\boldsymbol{Z} = \{\{ c_{n} \},\{ u_{n} \},r\}$. In (\ref{eq:variational_dist}), $\boldsymbol{Z}$ is divided into disjoint groups $Z_i,i = 1,\dots$, where each $Z_i$ is representing one of the hidden variables in $\boldsymbol{Z}$.  By maximizing the lower bound $\mathcal{L}(\mathbb{Q}\{\boldsymbol{Z} \})$, the variational distribution of each partition $\mathbb{Q}\{\boldsymbol{Z}_i\}$ is given by \cite[Chapter 10]{bishop2006pattern}:
\begin{equation}
\ln(\mathbb{Q}\{\boldsymbol{Z}_i\}) = \mathbb{E}_{j\neq i}\{ \ln( \mathbb{P} \{ \mathcal{A},\boldsymbol{Z}\} )\} + const,
\label{eq:general_update_rule}
\end{equation}
where $\mathbb{E}_{j\neq i}\{. \}$ is the expectation with respect to distributions $\mathbb{Q}\{\boldsymbol{Z}_j\}, j \neq i$. Then by plugging in $\mathbb{P} \{ \mathcal{A},\boldsymbol{Z}\} = \mathbb{P} \{ \mathcal{A},\boldsymbol{c},\boldsymbol{u},r\}$ from (\ref{eq:joint_dist}) and employing the exponential form of the distributions,  the variational distributions can be obtained. The constant value is determined by normalizing the distribution.

Using (\ref{eq:general_update_rule}), we can derive closed form expressions for parameters of the variational distributions. At each time step, after receiving the new observation vector,  $\boldsymbol{\alpha}^{(t)}$, the distribution of the hidden variables are updated using the derived update rules. Then, the updated distributions are used to concentrate the sensing energy on the more important coefficients of the signal. \par  

\section{Measurement Matrix Design}
\label{sec:matrix_design}
In this section, the distributions of the importance levels are exploited to design the measurement matrix at each time step $\boldsymbol{\Phi}^{(t)}$. The idea is to employ the information extracted from the previous measurements and focus the sensing energy on the ROI coefficients. \par 

In conventional compressive sensing methods, the sensing energy is distributed uniformly among the coefficients of the signal. In many standard methods, it is assumed that the column of the measurement matrix are scaled to be of unit norm. Thus, the total amount of sensing energy is $\| \Phi \|_F^2 = N$. In this work, we also assume that the available sensing energy is $N$. A constraint on the available sensing energy is necessary for any practical implementation. Also, without the constraint, the issue of noise would be irrelevant. \par 

In adaptive sensing procedures \cite{Malloy2014Near-OptimalSensing,Braun2015Info-GreedySensing,Haupt2012SequentiallySensing,Iwen2012AdaptiveEnvironments}, no energy is allocated to the coefficients that are not likely to be in support of the signal, i.e., ROI. However, in our problem, since we are dealing with time-varying signals, such hard decisions should be avoided. \par 

The key aspect of the proposed method is the allocation of sensing energy across the coefficients of the signal. In Section \ref{sec:Bayesian}, a Bayesian framework is introduced to obtain the distribution of the importance of each coefficient. Specifically, the norm of the $n^{\text{th}}$ column of the measurement matrix $\boldsymbol{\Phi^{(t)}}$ is given as:
\begin{equation}
\label{eq:col_norm}
\gamma_n^{(t)} = \sqrt{N}\frac{\bar{c}_{n}}{\eta}
\end{equation}
where $\bar{c}_{n}$ is the expected value of the importance level  of the $n^{\text{th}}$ coefficient of the signal, i.e. $\bar{c}_{n} = \mathbb{E}_{\mathbb{Q}\{ c_{n}\}}\{ c_{n}\}$. and $\eta$ is a constant to ensure that the energy constraint is met. Specifically, for $\eta = \sqrt{\sum_n \bar{c}_{n}^2}$, we will have $\| \Phi \|_F^2 = N$. \par 

Thus, at each time step the estimate of the signal is used to update the distribution of the hidden variables. Then, the inferred importance levels are exploited to tune the energy allocated to each coefficient of the signal.


\section{Numerical Experiments}\label{sec:results}
In this section, a series of numerical experiments are presented to highlight the performance gain of the ANCS. The primary performance metric used in our studies is time averaged normalized MSE (TNMSE), which is defined as
$
\frac{1}{T}\sum_{t = 1}^{T}\frac{\| \boldsymbol{x}^{(t)} - \hat{\boldsymbol{x}}^{(t)} \|_2^2}{\| \boldsymbol{x}^{(t)} \|_2^2}.
$
where $T$ is the number of time slots of the signal, $\| . \|_2$ is the $\ell_2$-norm of a vector, and $\hat{\boldsymbol{x}}^{(t)}$ is the estimate of $\boldsymbol{x}^{(t)}$ at time $t$. \par 

The parameters of the algorithm are set as follows. Since, no prior information is assumed on the importance levels of the coefficients, the parameters are initialized as $\beta^{1}_{n} = 1 = \beta^{0}_{n} = 1, ~\forall n$. This choice of parameters results in a uniform distribution for the importance levels. To initialize $b^1$ and $b^0$, it is reasonable to assume that at least half of the measurements are reliable. In our numerical experiments, we initialized $b^1 = 3$ and $b^0 = 1$, which means on average $75\%$ of the measurements are trustworthy. The maximum number of iterations for the inference algorithm is set to $40$, with possibility of early termination if $\frac{\sum_n(\bar{c}_n^{(k)} - \bar{c}_n^{(k-1)} )^2}{\sum_n(\bar{c}_n^{(k-1)})^2} \leq 10^{-6}$ at $k^{\text{th}}$ iteration. Moreover, a window length of $W = 5$ is used. \par 

In all the simulations to construct the measurement matrices, elements of the matrix were drawn from an i.i.d zero mean Gaussian distribution. For uniform sampling, the columns of the matrices are scaled to have unit norm. On the other hand, for ANCS, (\ref{eq:col_norm}) is used to realize the non-uniform distribution of energy among the columns. The total sensing energy of all the methods is assumed to be the same, i.e. $\| \boldsymbol{\Phi}^{(t)} \|_F^2 = N, \ \forall t$.\par 

As a performance benchmark and to quantify the performance improvement obtained by the ANCS, we exploit the proposed method as the sampling step of an $\ell_1$ minimization recovery algorithm. Specifically, the estimate of the signal is obtained by solving an $\ell_1$ minimization problem, given by: 
$$
\hat{\boldsymbol{x}}^{(t)} = \arg \min \| \boldsymbol{x} \|_1 , \text{ s.t. } \| \boldsymbol{y}^{(t)} - \boldsymbol{\Phi}^{(t)} \boldsymbol{x} \| _2 \leq c,
$$
where $\|\boldsymbol{x} \|_1=\sum_n|x_n|$ and $c$ is set to be equal to $\sigma_n\sqrt{M}$. To solve the problem, CVX  \cite{Grant2008GraphPrograms,Grant2014CVX:2.1}, which is a toolbox for specifying and solving convex problems, is used.

\subsection{Performance evaluation for sparse signals in canonical basis}
\label{subsec:results_canonical}
For the first experiment, the performance gain of ANCS is quantified for the signals that are sparse in canonical basis. To model the temporal correlation, both in amplitude and support of the signal, the signal is assumed to be outcome of two random processes. Specifically, a binary vector $\boldsymbol{s}^{(t)} = [s_1^{(t)},\dots,s_N^{(t)}]^T$ describes the support of the signal at time $t$. $s_n = 1$ indicates the coefficients in the support and $s_n^{(t)} = 0$ denotes the zero coefficients. Coefficients of $\boldsymbol{s}^{(t)}$ are assumed to be independent and a Markov chain process is defined for each of the coefficients. The Markov chain processes are described by $p_{01} = \mathbb{P} \{s_n^{(t)} = 1 |  s_n^{(t-1)} = 0 \}$ and $\lambda = \mathbb{P} \{s_n^{(t)} = 1\}, \  \forall n,t $. Thus, $\lambda$ is related to the sparsity level of signal. \par 

Furthermore, a second process models the amplitude of the large coefficients. We employ an independent Gauss-Markov process for each of the coefficients of the signal. Amplitude of the $n^{\text{th}}$ coefficient evolves over time as:
$a_n^{(t)} = (1 - \rho)a_n^{(t-1)} + 
\rho \nu_n^{(t)}.
$
Here, $\rho$ is a constant between $0$ and $1$ and controls the degree of correlation. For $\rho = 1$, the amplitude would be an uncorrelated Gaussian random process. $\nu_n^{(t)}$ is the amount of variation among two consecutive time steps and is modeled with $\mathcal{N}(0,\sigma_{L}^2)$. Thus the mean of the process is assumed to be $0$. At each time step, the coefficients of the signal are constructed as $x_n^{(t)} = a_n^{(t)}s_n^{(t)}. $\par

\begin{figure}
\centering     
\subfigure[ ]{
\includegraphics[width=1.5in,angle=0]{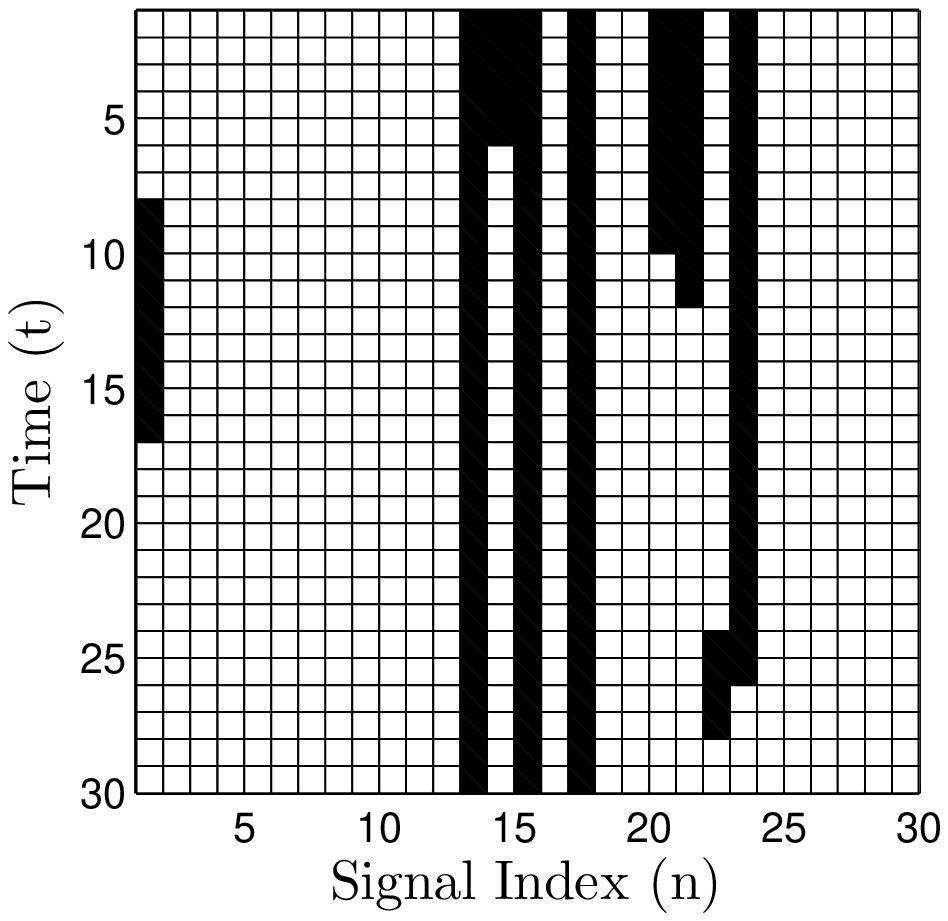}\vspace{-1.5mm}
\label{subfig:supportvstime}
}
\subfigure[ ]{
\includegraphics[width=1.7in,angle=0]{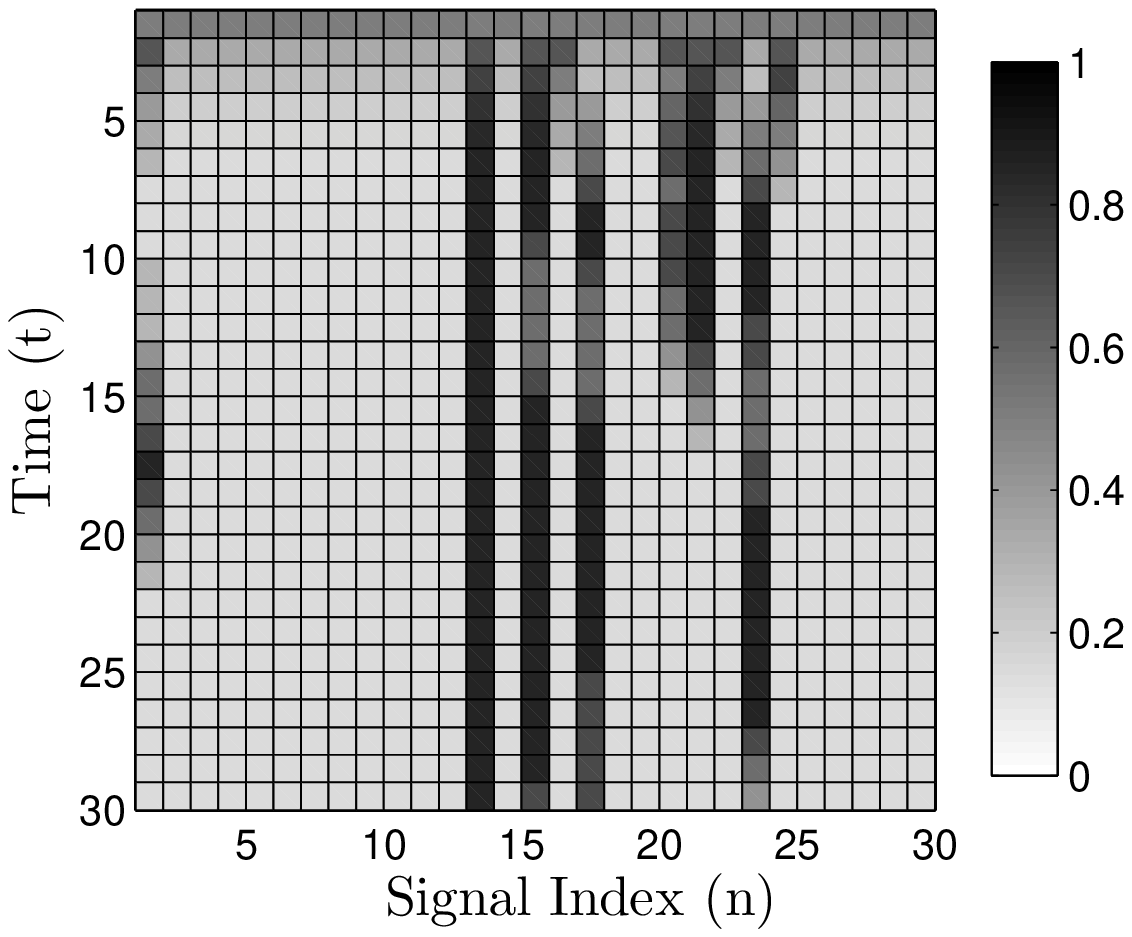}\vspace{-1.5mm}
\label{subfig:importancevstime}
}
\caption{ \small   \subref{subfig:supportvstime} Support of the signal and \subref{subfig:importancevstime} the expected value of the inferred importance levels, i.e., $\bar{c}_{n}$, for the first $30$ coefficients of the signal. $M = 60$, $N = 200$, SNR $= 20$ dB, and $W = 5$. }
\label{fig:importanceovertime}
\vspace{-5mm}
\end{figure}

The simulation parameters are set as follows, unless otherwise is stated. We assume that the signal of interest is of length $N = 200$ with sparsity level of $\lambda = \frac{K}{N} = 0.1$. The variance of noise, i.e., $\sigma_n^2$, is set to have a signal-to-noise ratio (SNR) of $20$ dB. Other model parameters are set as $\rho = 0.2$, $p_{01} = 0.02$, $\sigma_L = 10$, and $T = 30$. \par 

To detect the ROI, i.e., support of the signal, after determining the estimate of the signal $\hat{\boldsymbol{x}}^{(t)}$, a simple thresholding is performed. Specifically, $\alpha_n^{(t)}$ is set to $1$, if $\hat{x}_n^{(t)} \geq 1$. 

Figure \ref{fig:importanceovertime} shows the evolution of the inferred importance levels, i.e., $\bar{c}_n$, over time for $n = 1,2, \dots,30$. In other words, this figure illustrates how the sensing energy is distributed among the coefficients at each time step. As it is clear, at the first time step, $\bar{c}_n = 0.5, \forall n$, indicating unbiased estimate of importance levels when no further information is available. However, as more measurements are received, uncertainty decreases and the support of the signal is revealed. It is also worthwhile to point out that an error in the ROI detection procedure can potentially impact up to $W = 5$ time slots. Error propagation, as well as computational complexity, are the main reasons that choosing large values for $W$ should be avoided. \par 



\begin{figure}
\centering
\includegraphics[width=2.6in,angle=0]{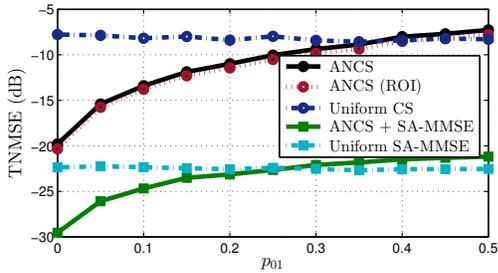}\vspace{-1.5mm}
\caption{\small Performance of ANCS for different values of $p_{01}$. The total sensing energy is the same for all the methods. $N = 200$, SNR $= 20$ dB, and $T = 30$, and $M = 60$. }
\label{fig:TNMSEvsP}
\vspace{-5mm}
\end{figure}

To study the performance of ANCS for different levels of temporal correlation, Figure \ref{fig:TNMSEvsP} illustrates the TNMSE of ANCS for different values of $p_{01}$. The results are averaged over substantially large number of Monte-Carlo Trials. Here, ANCS is employed also as the sampling step of Support-aware MMSE, as well as the $\ell_1$ minimization recovery method. SA-MMSE calculates the minimum mean square error estimate of the signal when the support of the sparse signal is known. The actual support of the signal, $\sigma_L^2$, $\sigma_n^2$, and $\rho$ are provided as the inputs of the SA-MMSE algorithm. The performance of SA-MMSE is an indicator of lowest MSE achievable by a recovery algorithm.\par 

For small values of $p_{01}$, the signal is nearly static over time. Thus, the method is able to detect the support accurately and the TNMSE is decreased significantly. Furthermore, since the signal is sparse in canonical basis and the support of the signal is set to be the ROI, overall recovery error is the same as the recovery error of the ROI coefficients. It is due to the fact that whole energy of the signal is concentrated in the ROI. As it can be noticed in the figure, for $p_{01} = 0$, ANCS can enhance the performance of the $\ell_1$ minimization algorithm substantially. 
As $p_{01}$ increases, the support of the signal changes over time and the observations of previous time steps become less informative about the signal and the performance gain of ANCS decreases. However, for values of $p_{01} < 0.3$, nonuniform recovery of the signal is still achieved. \par 

\begin{figure}
\centering     
\includegraphics[width=2.6in,angle=0]{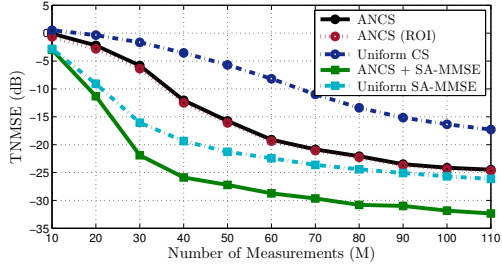}\vspace{-1.5mm}
\caption{ \small  TNMSE (in dB) for
of different recovery algorithm with and without ANCS as the sampling step for $N = 200$, SNR $= 20$ dB, and $T = 30$. }
\label{fig:ErrorvsM}
\vspace{-5mm}
\end{figure}

Figure \ref{fig:ErrorvsM} compares the performance of different recovery algorithms with uniform sampling and ANCS as the sampling step for different number of measurements. As it is clear, ANCS can decrease the TNMSE up to $7$ dB, compared to $\ell_1$ minimization recovery, and can reduce the required number of measurements. As an example, to achieve a TNMSE of $-15$ dB, ANCS employs about $50\%$ of the measurements required by uniform sampling, highlighting one of the major benefits of ANCS: for a sparse signal in canonical basis, ANCS is able to reduce the recovery error and number of required measurements substantially.\par

\begin{figure}
\centering     
\includegraphics[width=2.6in,angle=0]{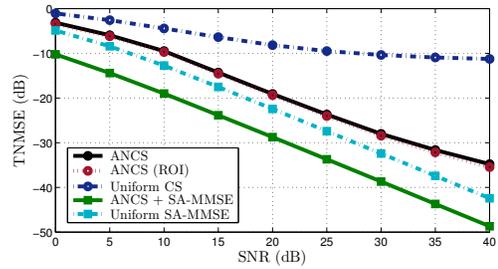}\vspace{-1.5mm}
\caption{ \small Performance of $\ell_1$ minimization and SA-MMSE with and without ANCS as the sampling step in terms of TNMSE (in dB). of different methods for $N = 200$, $p_{01} = 0.02$, and $M = 60$. }
\label{fig:ErrorvsSNR}
\vspace{-5mm}
\end{figure}

To highlight the performance gain achieved by ANCS in low SNR regimes, Figure \ref{fig:ErrorvsSNR} depicts TNMSE of different methods versus SNR. It is easy to notice that the performance of ANCS is very close to SA-MMSE with uniform sampling, which is MSE-optimal. This is one of the main benefits of adaptive CS. As mentioned in Section \ref{sec:intro}, it is known that adaptive CS provides the opportunity to detect and estimate signals at lower SNRs. Furthermore, performance of SA-MMSE algorithm in Figure \ref{fig:ErrorvsM} and Figure \ref{fig:ErrorvsSNR} illustrates that ANCS is able reduce the lower bound of recovery error by up to $6$ dB. \par 

\subsection{Performance evaluation for sparse signals in the DCT domain}
\label{subsec:DCT}
In this series of experiments, the performance gain achieved by ANCS is evaluated for signals that are not sparse in canonical basis, but has a sparse representation in some proper domain. In our numerical experiments, we employed DCT domain as the sparsifying basis. \par 

To generate the sparse signal in DCT domain, the same procedure explained in Section \ref{subsec:results_canonical} is exploited. Specifically, let $\boldsymbol{u}^{(t)} = \boldsymbol{\Psi}\boldsymbol{x}^{(t)}$ represent the sparse representation of the signal of interest, $\boldsymbol{x}^{(t)}$, in DCT domain. $\boldsymbol{\Psi}$ denotes the DCT transform matrix. To generate a time correlated signal, elements of $\boldsymbol{u}^{(t)}$ are constructed as $u_n^{(t)} = s_n^{(t)} a_n^{(t)}$, where $s_n^{(t)}$ and $a_n^{(t)}$ are outcome of two random processes described in Section \ref{subsec:results_canonical}. To reconstruct the signal, we use $\hat{\boldsymbol{x}}^{(t)} = \boldsymbol{\Psi}^T\hat{\boldsymbol{u}}^{(t)}$, where
$$
\hat{\boldsymbol{u}}^{(t)} = \arg \min \| \boldsymbol{u} \|_1 , \text{ s.t. } \| \boldsymbol{y}^{(t)} - \boldsymbol{\Xi}^{(t)} \boldsymbol{u} \| _2 < c,
$$
and $\boldsymbol{\Xi}^{(t)} = \boldsymbol{\Phi}^{(t)}\boldsymbol{\Psi}^T$.\par 

Furthermore, to model the variation of ROI over time, a new set of binary Markov processes is employed. This means that the probability of a coefficient being in the ROI is independent from its location and its value. To describe this Markov process, for simplicity, we use the same set of parameters as the random process corresponding to the support of the signal, i.e., $\lambda$ and $p_{01}$. Hence, the rates of change for support of $\boldsymbol{u}^{(t)}$ and the ROI in $\boldsymbol{x}^{(t)}$ are assumed to be the same. It is also assumed that the ROI detection algorithm may report erroneous observations to ANCS. \par

\begin{figure}
\centering
\includegraphics[width=2.6in,angle=0]{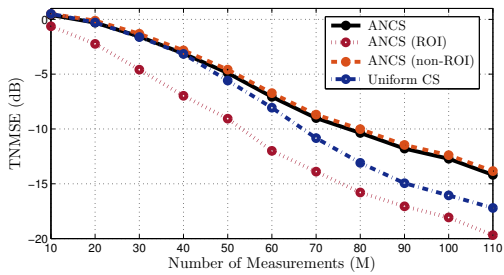}\vspace{-1.5mm}
 \caption{\small TNMSE (in dB) versus $M$ of ANCS. $N = 200$, SNR $= 20$ dB, and $T = 30$, and $M = 60$. }
\label{fig:DCTvs_M}
\vspace{-5mm}
\end{figure}

In Figure \ref{fig:DCTvs_M}, we evaluate the performance of ANCS versus the number of measurements $M$ for fault rate of $10\%$. This experiment also shows that the proposed ANCS is able to decrease the error of ROI coefficients up to $3$-$4$ dB for different number of measurements. This benefit comes at the cost of losing performance on total recovery error. Interestingly, for smaller values of $M$, this benefit comes at almost no cost and without losing any performance for non-ROI entries. \par 

\begin{figure}
\centering
\includegraphics[width=2.6in,angle=0]{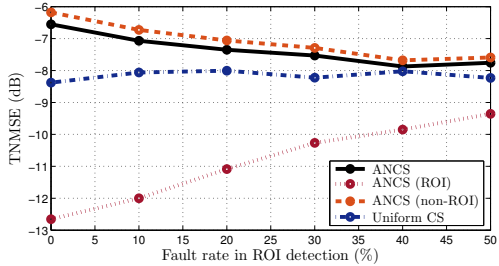}\vspace{-1.5mm}
 \caption{\small Plot of TNMSE (in dB) of ANCS for different values of fault rate. 
 $N = 200$, SNR $= 20$ dB, and $T = 30$, and $M = 60$. }
\label{fig:DCT_TNMSEvsMICmodel}
\vspace{-3mm}
\end{figure}

Finally, as it was expected, Figure \ref{fig:DCT_TNMSEvsMICmodel} illustrates that as ANCS receives more faulty data from the ROI detection algorithm, its performance becomes more similar to conventional CS with uniform sampling. This is because the faulty data prevents the inference algorithm from gaining certainty on the location of ROI coefficients. However, even for fault rates of as much as $50\%$, non-uniform recovery of the signal is achieved. \par 

\section{conclusions}\label{sec:conclusions}
This work presented \emph{adaptive non-uniform compressive sampling} (ANCS) for time-varying sparse signals.
The main idea is to employ the observations of previous time slots to infer the region of interest (ROI) in the signal and concentrate the sensing energy on the corresponding coefficients. 
For that, we presented a Bayesian framework, by modeling the overall and coefficient-specific reliability of the ROI detection algorithm. \par 

The results show that the proposed framework is able to achieve the desired non-uniform recovery and can decrease the error in ROI significantly for signals that are sparse or have a sparse representation in a proper basis. The results also illustrated that the proposed method is particularly advantageous for signals that are sparse in canonical basis. For such signals, ANCS results in substantial improvement in accuracy of estimation. \par 


\section{Acknowledgments}
This material is based upon work supported by the National Science Foundation under Grant No. ECCS-1418710.

\footnotesize{
\balance
\bibliography {Mendeley} 

\begin{thebibliography}{10}

\bibitem{Donoho2006CompressedSensing}
D.~Donoho, ``{Compressed sensing},'' {\em IEEE Transactions on Information
  Theory}, vol.~52, pp.~1289--1306, 4 2006.

\bibitem{Candes2006CompressiveSampling}
E.~Cand{\`{e}}s, ``{Compressive sampling},'' {\em Proceedings oh the
  International Congress of}, 2006.

\bibitem{shahrasbi2016model}
B.~Shahrasbi and N.~Rahnavard, ``{Model-Based Nonuniform Compressive Sampling
  and Recovery of Natural Images Utilizing a Wavelet-Domain Universal Hidden
  Markov Model},'' {\em IEEE Transactions on Signal Processing}, vol.~PP,
  no.~99, p.~1, 2016.

\bibitem{Uddin2011PhotoNet:Awareness}
M.~Y.~S. Uddin, H.~Wang, F.~Saremi, G.-J. Qi, T.~Abdelzaher, and T.~Huang,
  ``{PhotoNet: A Similarity-Aware Picture Delivery Service for Situation
  Awareness},'' in {\em 2011 IEEE 32nd Real-Time Systems Symposium},
  pp.~317--326, IEEE, 11 2011.

\bibitem{Rahimpour2016DistributedNetworks}
A.~Rahimpour, A.~Taalimi, J.~Luo, and H.~Qi, ``{Distributed object recognition
  in smart camera networks},'' in {\em 2016 IEEE International Conference on
  Image Processing (ICIP)}, pp.~669--673, IEEE, 9 2016.

\bibitem{Leinonen2015SequentialNetworks}
M.~Leinonen, M.~Codreanu, and M.~Juntti, ``{Sequential Compressed Sensing With
  Progressive Signal Reconstruction in Wireless Sensor Networks},'' {\em IEEE
  Transactions on Wireless Communications}, vol.~14, pp.~1622--1635, 3 2015.

\bibitem{Sani2016DistributedNetworks}
A.~Sani and A.~Vosoughi, ``{Distributed Vector Estimation for Power- and
  Bandwidth-Constrained Wireless Sensor Networks},'' {\em IEEE Transactions on
  Signal Processing}, vol.~64, pp.~3879--3894, 8 2016.

\bibitem{Rahmani2016ASampling}
M.~Rahmani and G.~Atia, ``{A Subspace Learning Approach for High Dimensional
  Matrix Decomposition with Efficient Column/Row Sampling},'' 2016.

\bibitem{Hosseini2016Cloud-basedPrediction}
M.-P. Hosseini, H.~Soltanian-Zadeh, K.~Elisevich, and D.~Pompili,
  ``{Cloud-based Deep Learning of Big EEG Data for Epileptic Seizure
  Prediction},'' in {\em Proc. of IEEE Global Conference on Signal and
  Information Processing (GlobalSIP)}, (Greater Washington, D.C., USA), IEEE,
  12 2016.

\bibitem{Fragkiadakis2014AdaptiveApplications}
A.~Fragkiadakis, P.~Charalampidis, and E.~Tragos, ``{Adaptive compressive
  sensing for energy efficient smart objects in IoT applications},'' in {\em
  2014 4th International Conference on Wireless Communications, Vehicular
  Technology, Information Theory and Aerospace {\&} Electronic Systems
  (VITAE)}, pp.~1--5, IEEE, 5 2014.

\bibitem{Ziniel2013DynamicPassing}
J.~Ziniel and P.~Schniter, ``{Dynamic Compressive Sensing of Time-Varying
  Signals Via Approximate Message Passing},'' {\em IEEE Transactions on Signal
  Processing}, vol.~61, pp.~5270--5284, 11 2013.

\bibitem{Wijewardhana2016AMeasurements}
U.~L. Wijewardhana and M.~Codreanu, ``{A Bayesian Approach for Online Recovery
  of Streaming Signals from Compressive Measurements},'' {\em IEEE Transactions
  on Signal Processing}, pp.~1--1, 2016.

\bibitem{Shahrasbi2011TC-CSBP:Propagation}
B.~Shahrasbi, A.~Talari, and N.~Rahnavard, ``{TC-CSBP: Compressive sensing for
  time-correlated data based on belief propagation},'' in {\em 2011 45th Annual
  Conference on Information Sciences and Systems}, pp.~1--6, IEEE, 3 2011.

\bibitem{Malloy2014Near-OptimalSensing}
M.~L. Malloy and R.~D. Nowak, ``{Near-Optimal Adaptive Compressed Sensing},''
  {\em IEEE Transactions on Information Theory}, vol.~60, pp.~4001--4012, 7
  2014.

\bibitem{Braun2015Info-GreedySensing}
G.~Braun, S.~Pokutta, and Y.~Xie, ``{Info-Greedy Sequential Adaptive Compressed
  Sensing},'' {\em IEEE Journal of Selected Topics in Signal Processing},
  vol.~9, pp.~601--611, 6 2015.

\bibitem{Haupt2012SequentiallySensing}
J.~Haupt, R.~Baraniuk, R.~Castro, and R.~Nowak, ``{Sequentially designed
  compressed sensing},'' in {\em 2012 IEEE Statistical Signal Processing
  Workshop (SSP)}, pp.~401--404, IEEE, 8 2012.

\bibitem{Jordan2008graphical}
M.~J. Wainwright and M.~I. Jordan, ``{Graphical models, exponential families,
  and variational inference},'' {\em Foundations and
  Trends{\{}{\textregistered}{\}} in Machine Learning}, vol.~1, no.~1-2,
  pp.~1--305, 2008.

\bibitem{jordan1999variational}
M.~I. Jordan, Z.~Ghahramani, T.~S. Jaakkola, and L.~K. Saul, ``{An introduction
  to variational methods for graphical models},'' {\em Machine learning},
  vol.~37, no.~2, pp.~183--233, 1999.

\bibitem{Babagholami-Mohamadabadi2014AEstimation}
B.~Babagholami-Mohamadabadi, A.~Jourabloo, A.~Zarghami, and S.~Kasaei, ``{A
  Bayesian Framework for Sparse Representation-Based 3-D Human Pose
  Estimation},'' {\em IEEE Signal Processing Letters}, vol.~21, pp.~297--300, 3
  2014.

\bibitem{bishop2006pattern}
C.~M. Bishop, {\em {Pattern recognition and machine learning}}.
\newblock springer, 2006.

\bibitem{Iwen2012AdaptiveEnvironments}
M.~A. Iwen and A.~H. Tewfik, ``{Adaptive Strategies for Target Detection and
  Localization in Noisy Environments},'' {\em IEEE Transactions on Signal
  Processing}, vol.~60, pp.~2344--2353, 5 2012.

\bibitem{Grant2008GraphPrograms}
M.~Grant and S.~Boyd, ``{Graph implementations for nonsmooth convex
  programs},'' in {\em Recent Advances in Learning and Control} (V.~Blondel,
  S.~Boyd, and H.~Kimura, eds.), Lecture Notes in Control and Information
  Sciences, pp.~95--110, Springer-Verlag Limited, 2008.

\bibitem{Grant2014CVX:2.1}
M.~Grant and S.~Boyd, ``{CVX: Matlab Software for Disciplined Convex
  Programming, version 2.1}.'' http://cvxr.com/cvx, 3 2014.

\end{thebibliography}
\bibliographystyle{ieeetr}
}

\end{document}